\documentclass{elsart3}
\usepackage{graphicx}
\begin{document}

\begin{frontmatter}
\title{$\beta$-Detected NMR of $^8$Li in  the Normal State of 2H-NbSe$_2$}
\author[UBCPHYS]{D.Wang}
\author[UBCPHYS]{M.D. Hossain}
\author[TRIUMF]{Z. Salman}
\author[TRIUMF]{D. Arseneau}
\author[UOFA]{K.H. Chow}
\author[TRIUMF]{S. Daviel}
\author[UBCPHYS]{T.A.Keeler}
\author[UBCPHYS]{R.F. Kiefl\corauthref{cor}}
\corauth[cor]{Tel. 604 222 7511,
Fax: 604 222-1074,
email: kiefl@triumf.ca}  
\author[TRIUMF]{S.R. Kreitzman}
\author[TRIUMF]{C.D.P. Levy}
\author[TRIUMF]{G.D. Morris}  
\author[TRIUMF]{R.I. Miller}  
\author[UBCCHEM]{W.A. MacFarlane}
\author[UBCCHEM]{T.J. Parolin}
\and \author[UBCPHYS]{H.Saadaoui}
\address[TRIUMF]{TRIUMF, 4004 Wesbrook Mall, University of B.C., Vancouver, Canada V6T2A3}
\address[UBCPHYS]{Department of Physics and Astronomy, University of B.C., Vancouver, V6T1Z1}
\address[UOFA]{Department of Physics, University of Alberta, Edmonton Alberta, Canada T6G2J1} 
\address[UBCCHEM]{Department of Chemistry,  University of B.C.,  Vancouver, Canada V6T1Z1}

\begin{abstract}
$\beta$-NMR of isolated $^8$Li has been investigated in the normal state of 2H-NbSe$_2$.  In a high magnetic field of 3T  a single  resonance is observed with a  Gaussian line width of 3.5 kHz. The line shape varies  weakly as function of magnetic field and temperature but has a strong orientation dependence. The  nuclear electric  quadrupole splitting is unresolved implying that the electric field gradients are 10-100 times smaller than in other non-cubic crystals. The nuclear spin relaxation rate is also anomalously small but varies linearly with  temperature as expected for Korringa relaxation in a metal. These results suggest that Li adopts an interstitial position  between the weakly coupled  NbSe$_2$ layers and away from the conduction band.
\end{abstract}

\end{frontmatter}


Transition metal dichalcogenides such as NbSe$_2$ have a variety of interesting properties related to their layered crystal structure. In particular these materials  consist of layers of atoms which are metallically and covalently  bonded  within the layer but experience weak Van der Waals interactions with adjacent layers of atoms.  This leads to strong anisotropy in  both  electronic and mechanical properties. For example NbSe$_2$  shows a superconducting transition  at $T_c$=7.0K with a highly anisotropic penetration depth, coherence length and effective mass\cite{poole}. Mechanically it cleaves easily between the  NbSe$_2$ layers. In addition  similar compounds have also  been  examined as possible battery electrode materials since alkali atoms such as Li can be intercalated into the spaces  between the layers\cite{whittingham}. NMR studies on  Li compounds  such as LiMX$_2$ (M=Ti,V,Cr and X=S,Se) find that Li occupies octahedral sites surrounded by six X atoms in the Van der Waals gap\cite{prigge}.
  
In this paper we present a study of isolated $^8$Li in the normal state of 2H-NbSe$_2$ using the technique of $\beta$-detected NMR. $^8$Li is a radioactive spin 2 nucleus with a small electric quadrupole moment of +33 mB. The nuclear polarization can be monitored through beta decay as described below. In non-cubic materials this quadrupole moment couples to the electric field gradient giving rise to an electric  quadrupole  splitting of the resonance which is  typically in the range 10-100 kHz \cite{salman}. Surprisingly in NbSe$_2$  we observe a large single resonance centered at the Larmor frequency with a width typical of nuclear dipolar broadening\cite{prigge}.  This implies that the electric field gradient at the Li site in NbSe$_2$ is unusually small. The resonance varies weakly as a function of the temperature and magnetic field but has a strong orientation dependence. 
In high magnetic fields the spin relaxation rate varies linearly with temperature as expected for Korringa relaxation due to scattering with the conduction electrons. These results suggest that below room temperature  the implanted $^8$Li occupies an interstitial site between the NbSe$_2$ layers. The simplicity of the resonance suggests that $^8$Li could be used to probe the magnetic field distribution near the surface in the superconducting state.
Muon spin rotation has been used extensively in this regard as probe of  bulk superconductors \cite{sonier} and also near surfaces\cite{niedermayer}. 
 
The experiment was  performed at the ISAC radioactive ion beam facility  on the polarized beamline.  Details on the polarizer and spectrometers are given elsewhere\cite{kiefl,salman2,morris}. We mention here only that the highly polarized  (70\%) $^8$Li$^+$ ions were implanted into the sample with an energy of 28 keV corresponding to  a mean implantation depth of 150 nm. The beam intensity was about $10^6$s$^{-1}$ so that radiation damage is minimal.   
$^8$Li is the lightest nucleus suitable for $\beta$-detected NMR. This nuclear method has  the same basic principle as  muon spin rotation since the nuclear polarization is monitored  through the anisotropic $\beta$ decay of a polarized radioactive nucleus. The nuclear resonance in a static magnetic field, $H_0$ can be detected by measuring the time averaged  nuclear polarization with a continuous beam as function of a small perpendicular RF magnetic field.  The position and shape of the resonance(s) is sensitive  to  the local electronic and magnetic environment. Alternatively one may introduce a short pulse of $^8$Li  and measure the time evolution of the polarization analogous to  muon spin rotation at a pulsed facility. 
The sample in this experiment  was a single crystal of 2H-NbSe$_2$ about 4 mm is diameter and 0.1 mm thick attached to a sapphire plate. It was cleaved just prior to introducing it into the ultra high vacuum ($10^{-9}$ torr) to minimize contact with air. The beam was focused onto the sample so that there was no discernible background signal from the sapphire or elsewhere.


A typical resonance in a high magnetic field is shown in the top panel of Fig. 1.  Excellent fits were obtained to a Gaussian line shape. The fitted line width $\sigma$ defined as the FWHM equals 3.5(1)kHz  and is  weakly dependent on the RF power level, indicating the measured line width is close to the intrinsic width.  2H-NbSe$_2$ is hexagonal and consequently  one expects the resonance to be split by an electric quadrupolar interaction, present at a non-cubic site. The absence of a resolved splitting implies the electric field gradient at the Li site is 10-100 times  smaller than observed in most other non-cubic crystals\cite{salman}. The observed line width is attributed to the nuclear moments of $^{93}$Nb (100\% S=9/2) and $^{77}$Se (7.6\% S=1/2) plus the unresolved quadrupolar splitting.
The middle panel in Fig. 1 shows the same resonance in a much smaller static magnetic field of 10~mT. Note the line shape is almost identical. This confirms there is no large contribution to the line broadening in high field from a spread in Knight shifts or chemical shifts since in this case one would expect a decrease in the line width at low field. When the magnetic field is oriented perpendicular to the c-axis (see bottom resonance in Fig.1) the line width decreases  to 1.5kHz. This is consistent with Li occupying a site in the Van der Waals gap since then one would expect line broadening from  Nb moments and  quadrupolar splittings to be greatest when the field is parallel to the c-axis. For example any broadening due to unresolved  splittings arising from  a  quadrupolar interaction which is axially symmetric about the  c-axis  should vary  as  $1-3\cos^2\theta$ where $\theta$ is the angle between the magnetic field and c-axis. 
The observed orientation dependence is slightly larger than this simple model predicts,
suggesting a more detailed calculation including fluctuating nuclear dipolar fields is needed to explain the results quantitatively.

\begin{figure}
\centering
\includegraphics[width=\columnwidth]{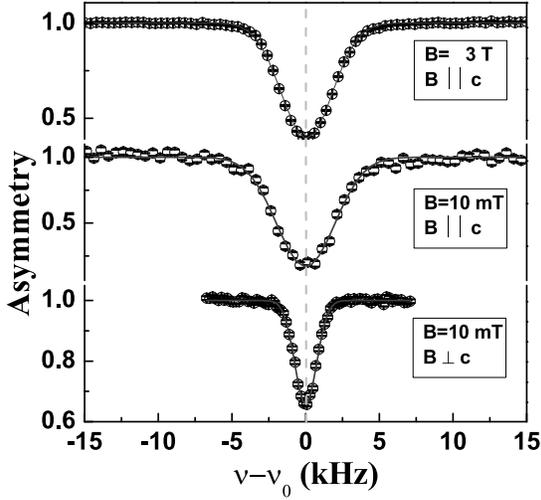}
\caption{The $\beta$-NMR resonance in NbSe$_2$ as a function frequency for different fields and orientations. The top two scans were taken with the field  parallel  to the c-axis  but at  two very different fields; whereas, the bottom scan is with the field  perpendicular to the c-axis. The temperature is 10K in all cases. 
}
\end{figure}

There is a weak temperature dependence to the resonance line shape as may be seen from Figs. 2.  In particular the resonance broadens slightly and becomes slightly asymmetric at higher temperatures.  This is also evident from the fits to a single Gaussian line shape (see Fig. 3) which show the average frequency and line width increase with temperature while the peak amplitude decreases. One explanation is that a second site at a slightly higher frequency becomes partially occupied at higher temperatures. The observed temperature dependence in the amplitude, frequency and line width would then arise from changes in the occupation probability  for the  two unresolved resonances.   In order to explain the observed effects  the second site would have to be shifted on the order a 1-2 kHz or 50-100 ppm. Similar thermally induced transitions between sites have been observed for $^8$Li in other materials such as Cu
\cite{fullgrabe}, Ag\cite{morris} and GaAs \cite{chow}.

\begin{figure}
\centering
\includegraphics[width=\columnwidth]{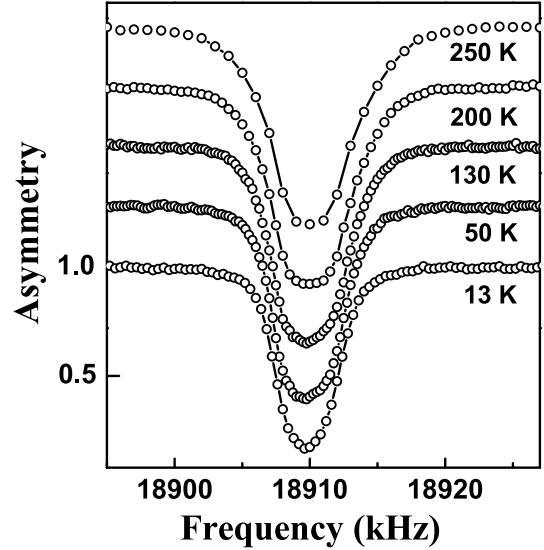}
\caption{Temperature variation of the resonance in a large applied field of 3T parallel to the c-axis. The RF power was about 35 times that used for the resonances in Fig. 1. Note the resonance broadens and becomes slightly asymmetric at higher temperatures. The curves shown simply connect the data points.}
\end{figure}

\begin{figure}
\centering
\includegraphics[width=\columnwidth]{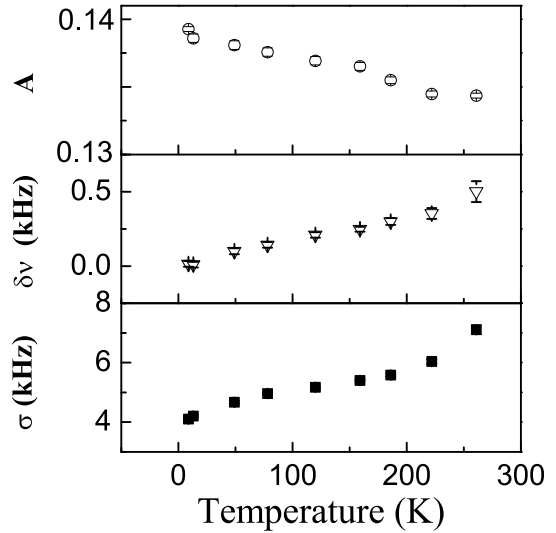}
\caption{The temperature dependence of the  peak amplitude (A), frequency shift($\delta \nu$) and  line width ($\sigma$), all  obtained from fits to a single Gaussian line shape.
}
\end{figure}

Measurements of the spin lattice relaxation rate were also carried out by introducing a short beam pulse of $^8$Li  into the sample and measuring the decay of the polarization as a function of time with no RF field. Good fits to the data were obtained by fitting to a single exponential as shown in Fig. 4a. The relaxation rates as a function of temperature are shown in Fig. 4b. The linear behaviour is characteristic of Korringa relaxation. The proportionality constant $9(1)\times 10^{-5}$ K$^{-1}$s$^{-1}$ is about 10 times smaller than  for  $^8$Li in a metal such as Ag\cite{morris}. This is consistent with Li in NbSe$_2$ occupying a site in the Van der Waals gap where the overlap with the conduction band is small. Note there is also a small relaxation at T=0K which we attribute to the residual effects of nuclear spin dynamics which completely dominate the Korringa  relaxation in low field  but are highly suppressed in 3T. We find no evidence of a discontinuity in $1/T_1$ or the resonance characteristics around 30K where there is  charge density wave transition\cite{skripov}. This is further evidence that the Li is weakly coupled to the conduction electron band as one would expect for a site in the Van der Waals gap. $^8$Li should therefore be an ideal probe for investigating the magnetic field distribution in the superconducting state near the surface.

\begin{figure}
\centering
\includegraphics[width=\columnwidth]{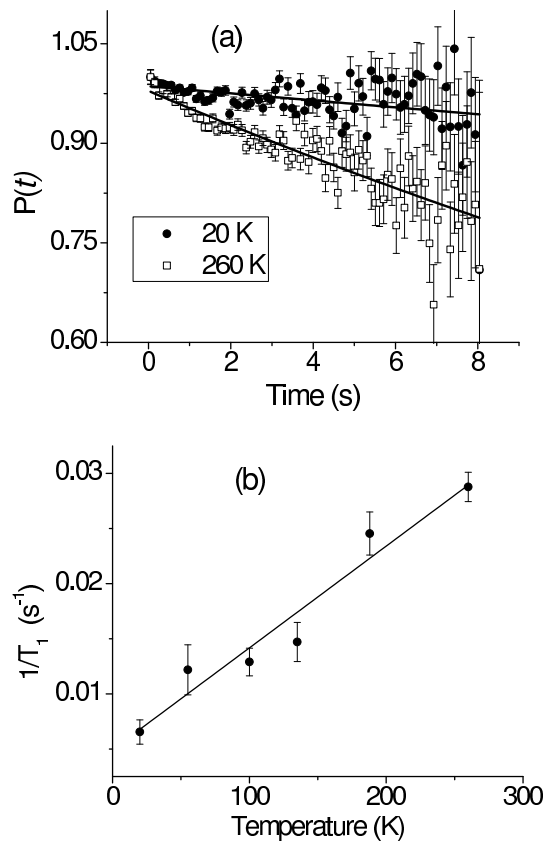}
\caption{(a) The time evolution of the spin polarization  $^8$Li in NbSe$_2$  in a magnetic field of 3T applied along the c-axis. (b)  The fitted spin relaxation rate as a function of temperature.
}
\end{figure}


In conclusion a single  $\beta$-NMR resonance centered at the Larmor frequency with no resolved quadrupolar splittings is observed in NbSe$_2$ from  10K up to RT. The resonance line shape is a weak function of magnetic field and temperature.  However the line width is more than a factor of two smaller with B perpendicular to c.
The Korringa relaxation rate is also very small. These results are consistent with Li occupying an octahedral site in the Van der Waals gap. The simplicity of the resonance suggests that $\beta$-NMR of $^8$Li is well suited to studies of the vortex lattice near the surface of NbSe$_2$. 

We would like to thank Joe Brill at the University of Kentucky for providing the sample.
This research was supported by the Center for Materials and Molecular Research at TRIUMF, the Natural Sciences and Engineering Research Council of Canada and the Canadian Institute for Advanced Research. We would especially like to thank Rahim Abasalti and  Bassam Hitti for their expert technical support.

\end{document}